\documentclass[mypaper,8pt,twoside]{CoAst}
\usepackage{epsf,graphicx,fancyhdr}
\renewcommand{\chaptermark}[1]{\markboth{#1}{}}

\newcommand{\apj}{ApJ}
\newcommand{\aap}{A\&A}  
\newcommand{\mnras}{MNRAS}

\newcommand{\dss}{$\delta$~Scuti stars}                     
\newcommand{\ds}{$\delta$~Scuti star}

\newcommand{\kopf}{\small\itshape Comm. in Asteroseismology\\ Vol. 149 (?), 2007}
\newcommand{\Authors}[1]{\begin{center}\normalsize\bf\sf #1 \end{center}}

\renewcommand{\author}[1]{\begin{center}\normalsize\bf\sf #1 \end{center}}
\newcommand{\Address}[1]{\begin{center}\small\sf #1 \end{center}}

\renewenvironment{abstract}{\section*{Abstract}\normalsize\sf}{}
\newcommand{\References}[1]{\begin{flushleft}{\large References\\}\vspace*{2mm}\small #1 \end{flushleft}}

\newcommand{\chapterDSSN}[2]{\chapter[\sf\normalsize #1\\ \footnotesize \hspace*{5mm}by #2 \sf\normalsize][]{#1\\}\rhead[\fancyplain{}{\sf\footnotesize \center{#1}}]{\fancyplain{}{\sffamily\thepage}}\lhead[\fancyplain{\kopf}{\sffamily\thepage}]{\fancyplain{\kopf}{\sf\footnotesize \center{#2}}}}

\newcommand{\figureDSSN}[5]{\begin{figure}[#4]
\centering
\includegraphics*[#5]{#1}
\caption{#2}
\label{#3}
\end{figure}}

\renewcommand{\chaptername}{}
\newcommand{\acknowledgments}[1]{\vspace*{5mm}\noindent\begin{bf}Acknowledgments. \end{bf} #1}

\newcommand{\simbad}{{\sc simbad}}

\newcommand{\publ}{$\surd$}    
\newcommand{\prep}{$\bigcirc$} 

\newcommand{\mhz}{$\mu$Hz}

\newcommand{\apjl}{ApJL}

 \newcommand{\nat}{Nature}

\begin{document}
\sf

\chapterDSSN{Asteroseismology with the WIRE satellite}{H.\ Bruntt}

\Authors{H.\ Bruntt}
\Address{School of Physics A28, University of Sydney, 2006 NSW, Australia}

\noindent
\begin{abstract}
I give a summary of results from the WIRE satellite, which has been used
to observe bright stars from 1999--2000 and 2003--2006.  The WIRE targets
are monitored for up to five weeks with a duty cycle of 30--40\%.
The aim has been to characterize the 
flux variation of stars across the Hertzsprung-Russell diagram. 
I present an overview of the results for solar-like stars, \dss, giant stars,
and eclipsing binaries.
\end{abstract}

\section{Introduction}

The Wide-field Infra-Red Explorer (WIRE) satellite was launched on 4 March
1999 with the aim to study star-burst galaxies (Hacking et al.\ 1999). The
mission was declared a failure only a few days after launch when it was
realised that the hydrogen coolant for the main camera had leaked.  Since
May 1999 the star tracker on board WIRE has been used to measure the
variability of bright stars (Buzasi et al.\ 2000).  Previous reviews of
the performance and science done with WIRE were given by Buzasi (2001,
2002, 2004), Laher et al.\ (2000), and Buzasi \& Bruntt (2005).

\section{Observing with WIRE}

WIRE is in a Sun-synchronous orbit with a period that has decreased from
$96$ to $93$ minutes from 1999 to 2006.  Constraints from the pointing of
the solar panels limits pointing in two roughly $\pm30^\circ$ strips
located perpendicular to the Sun-Earth line (Buzasi et al.\ 2000).  In
order to limit scattered light from the illuminated face of the Earth the
satellite switches between two targets during each orbit.  Each target has
a duty cycle of typically 30--40\%.

The star tracker has a $52$-mm aperture and a $512^2$ pixel SITe CCD.  
Windows of $8\times8$ pixels centered on the star are read out from the
CCD at a cadence of 10~Hz.  An example is shown in the left panel of
Fig.~1. During the first few months of operation only the primary target
was read out in the 10~Hz high cadence mode, but after refining the
on-board software up to five targets were read out (each target read out
at 2~Hz). In the right panel of Fig.~1 I show the distribution of $x,y$
positions for 56\,000 windows centered on the main target ($\alpha$~Cir).
The FWHM of the distribution is just one hundredth of a pixel. One pixel
on the CCD corresponds to about one arc minute. For details on the
photometric pipeline and a discussion of scattered light see Bruntt et
al.\ (2005).

In the early WIRE runs from 2000--1 the field was slowly rotating which
meant that the secondary targets moved across the CCD at timescales of one
pixel every few days (depending on the distance from the main target which
is centered on the CCD). This data is thus only of limited use since it is
not possible to take flat fields. Due to lack of funding WIRE was put into
sleep mode for about two years from September 2001 -- November 2003.  For
the past three years WIRE has observed in a new mode where the secondary
stars stay fixed on the same position on the CCD. As a consequence, the
number of stars observed with high photometric precision has increased
from a few dozen to more than two hundred.

\figureDSSN{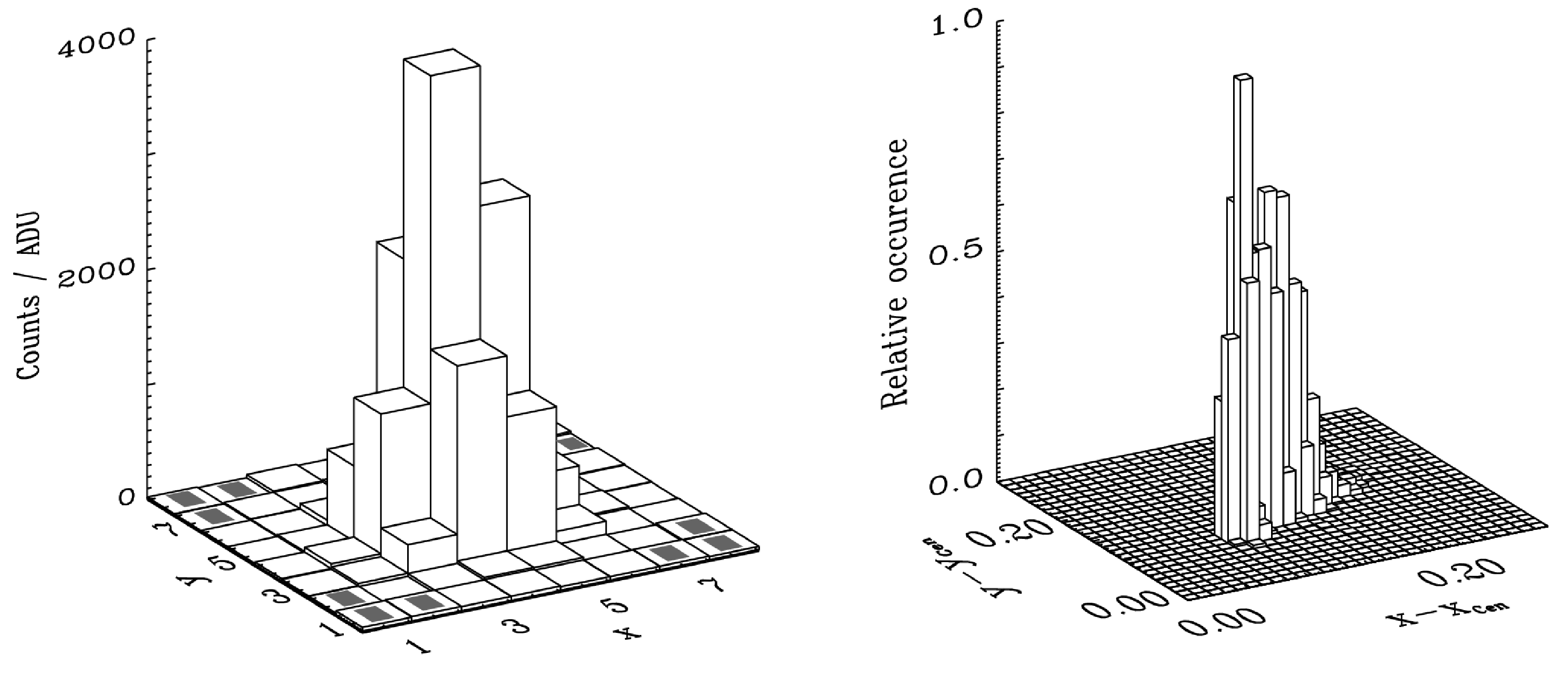}{The left panel shows a CCD window from WIRE. 
The grey boxes mark the pixels used for determination of the sky
background. The right panel shows the distribution of the $x,y$ position
of the central target from 56\,000 CCD windows.}{Figure 1}{!ht}{clip,angle=0,width=115mm}


\section{An overview of stars observed with WIRE}

In Table~1 I list the brightest stars observed with WIRE from March 1999
to June 2006. There are 45 main sequence stars (luminosity class IV-V) on
the left part of the table and 45 evolved stars on the right.  I give the
common name of each star (usually the Bayer designation), the Henry Draper
number, $V$ magnitude, and spectral class. This information was extracted
from the \simbad\ database. I have also marked the stars for which the
analysis has been published (marked with a \publ) and the stars that are
currently being analysed (marked with a \prep).

In Fig.~2 I show the location in the Hertzsprung-Russell diagram of 200
stars observed with WIRE. In the following I will briefly discuss the main
results for different classes of stars.

\figureDSSN{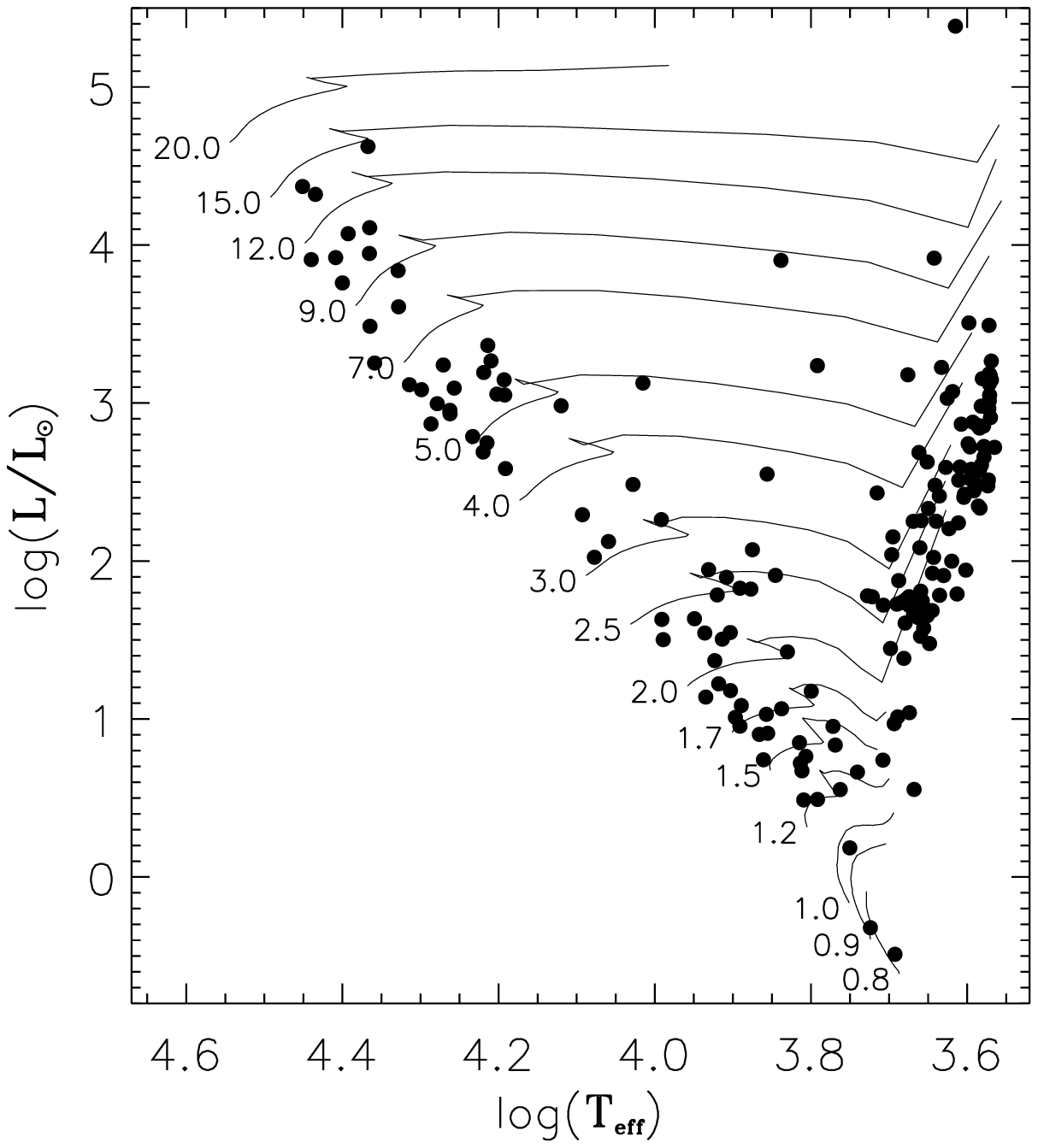}{Hertzsprung-Russell diagram of about 
200 stars observed with WIRE.}{Figure 2}{!ht}{clip,angle=0,width=100mm}

\begin{table}
\begin{center}
\caption{The brightest main sequence (left) and evolved stars (right) 
observed with WIRE.
The name, HD number, $V$ magnitude, and spectral type are given.
Stars whose observations have been published are marked by \publ, while
stars currently being analysed are marked by \prep.\label{tab:str}}
\setlength{\tabcolsep}{5pt} 
\begin{footnotesize}
\begin{tabular}{clrcl|clrcl}
\hline\hline
 & \multicolumn{1}{c}{Name} & \multicolumn{1}{c}{HD} & $V$ & \multicolumn{1}{c}{Type}&    
 & \multicolumn{1}{c}{Name} & \multicolumn{1}{c}{HD} & $V$ & \multicolumn{1}{c}{Type}\\
\hline                         
\publ &  $\alpha$ Cen      &128620 &  0.0 &     G2V   &  \publ &     $\alpha$ Boo     &124897 &  0.0 & K1.5III  \\
\publ &  $\alpha$ CMi      & 61421 &  0.3 &  F5IV-V   &        &         $\alpha$ Ori & 39801 &  0.6 &  M2Iab   \\
\publ &  $\alpha$ Aql      &187642 &  0.8 &     A7V   &  \publ & $\alpha$ UMa         & 95689 &  1.8 &  K0Iab   \\
      &       $\alpha$ Vir &116658 &  1.0 &B1III-IV   &        &       $\epsilon$ Car & 71129 &  2.0 &K3IIIva  \\
\publ &  $\beta$ Cru       &111123 &  1.3 &  B0.5IV   &        &          $\beta$ CMa & 44743 &  2.0 &B1II/III  \\
      &       $\alpha$ Leo & 87901 &  1.4 &     B7V   &  \prep &     $\alpha$ UMi     &  8890 &  2.0 &F7 Ib-II  \\
\prep &  $\lambda$ Sco     &158926 &  1.6 &B2IV+      &        &          $\beta$ UMi &131873 &  2.1 &   K4III  \\
\publ &  $\epsilon$ UMa    &112185 &  1.8 &  A0p      &        &         $\gamma$ Dra &164058 &  2.2 &   K5III  \\
\prep &  $\beta$ Aur       & 40183 &  1.9 &A2IV+      &        &         $\alpha$ Lup &129056 &  2.3 & B1.5III  \\
      &       $\alpha$ Pav &193924 &  1.9 &    B2IV   &  \publ & $\kappa$ Sco         &160578 &  2.4 & B1.5III  \\
      &       $\delta$ Vel & 74956 &  2.0 &     A1V   &        &       $\epsilon$ Peg &206778 &  2.4 &    K2Ib  \\
      &       $\gamma$ Leo & $-$   &  2.0 &      K0   &        &          $\beta$ Peg &217906 &  2.4 &M2.5II-I  \\
      &       $\sigma$ Sgr &175191 &  2.1 &     B2V   &        &         $\alpha$ Peg &218045 &  2.5 &   B9III  \\
      &        $\beta$ Leo &102647 &  2.1 &     A3V   &        &       $\epsilon$ Cyg &197989 &  2.5 &   K0III  \\
      &        $\beta$ Cas &   432 &  2.3 &    F2IV   &        &         $\gamma$ Aql &186791 &  2.7 &    K3II  \\
      &       $\delta$ Sco &143275 &  2.3 & B0.2IVe   &        &       $\epsilon$ Vir &113226 &  2.8 &   G8III  \\
      &         $\eta$ Cen &127972 &  2.3 & B1.5Vne   &        &           $\eta$ Peg &215182 &  2.9 &G2II-III  \\
      &       $\kappa$ Vel & 81188 &  2.5 &  B2IV-V   &        &        $\iota^1$ Sco &161471 &  3.0 &   F2Iae  \\
      &        $\zeta$ Oph &149757 &  2.6 &     O9V   &        &         $\alpha$ Ind &196171 &  3.1 &K0IIICNv  \\
      &       $\alpha$ Col & 37795 &  2.6 &   B7IVe   &        &          $\beta$ Col & 39425 &  3.1 &   K2III  \\
      &         $\eta$ Boo &121370 &  2.7 &    G0IV   &        &           $\phi$ Sgr &173300 &  3.2 &   B8III  \\
      &     $\upsilon$ Sco &158408 &  2.7 &    B2IV   &        &                G Sco &161892 &  3.2 &   K2III  \\
\publ &        $\beta$ Hyi &  2151 &  2.8 &    G2IV   &        &         $\kappa$ Oph &153210 &  3.2 &   K2III  \\
      &       $\alpha$ Ara &158427 &  2.8 &   B2Vne   &        &          $\beta$ Cep &205021 &  3.2 &B2IIIeva  \\
      &          $\pi$ Sco &143018 &  2.9 & B1V+      &        &        $\tau$ Sgr    &177716 &  3.3 &  K1IIIb  \\
      &        $\zeta$ Tau & 37202 &  3.0 &    B2IV   &        &       $\epsilon$ Cas & 11415 &  3.3 &   B3III  \\
\prep &       $\alpha$ Cir &128898 &  3.2 &     A5V   &        &          $\zeta$ Cep &210745 &  3.4 &K1.5Iab   \\
      &       $\delta$ UMa &106591 &  3.3 &     A3V   &  \publ & $\theta^2$ Tau       & 28319 &  3.4 &   A7III  \\
      &       $\delta$ Eri & 23249 &  3.5 &    K0IV   &        &            $\xi$ Hya &100407 &  3.5 &   G7III  \\
      &            $o$ Vel & 74195 &  3.6 &    B3IV   &        &         $\gamma$ Tau & 27371 &  3.7 &   K0III  \\
      &        $\beta$ Aql &188512 &  3.7 &    G8IV   &        &          $\beta$ Ind &198700 &  3.7 &    K1II  \\
      &     $\epsilon$ Eri & 22049 &  3.7 &     K2V   &        &            $\xi$ Dra &163588 &  3.7 &   K2III  \\
      &         $\rho$ Sco &142669 &  3.9 &  B2IV-V   &        &            $\nu$ Eri & 29248 &  3.9 &   B2III  \\
      &          $\pi$ Lup &133242 &  3.9 &     B5V   &        &          $\nu^2$ CMa & 47205 &  4.0 &K1III+    \\
\publ &         $\psi$ Cen &125473 &  4.0 &    A0IV   &        &       $\upsilon$ Boo &120477 &  4.1 & K5.5III  \\
      &          $\mu$ Eri & 30211 &  4.0 &    B5IV   &        &         $\delta$ Cep &213306 &  4.1 &  F5Iab   \\
      &         $\rho$ Lup &128345 &  4.0 &     B5V   &        &                  $-$ &  5848 &  4.2 &K2II-III  \\
      &          $\mu$ Ori & 40932 &  4.1 &     A2V   &        &                Q Sco &159433 &  4.3 &  K0IIIb  \\  
\publ &     $\epsilon$ Cep &211336 &  4.2 &    F0IV   &        &            $\pi$ Aur & 40239 &  4.3 &    M3II  \\    
      &             90 Tau & 29388 &  4.3 &     A6V   &        &               CE Tau & 36389 &  4.4 &  M2Iab   \\
      &            $o$ Lup &130807 &  4.3 &    B5IV   &        &             V761 Cen &125823 &  4.4 &B7IIIpva  \\  
      &       $\delta$ UMi &166205 &  4.3 &    A1Vn   &        &         $\sigma$ Lup &127381 &  4.4 &   B2III  \\    
      &       $\gamma$ Col & 40494 &  4.3 &  B2.5IV   &        &          $\nu^3$ CMa & 47442 &  4.4 &K0II/III  \\
      &       $\tau^2$ Lup &126354 &  4.4 &      F7   &        &           $\rho$ Cas &224014 &  4.5 &  G2Ia0e  \\
      &        $\iota$ Oph &152614 &  4.4 &     B8V   &        &               11 Cep &206952 &  4.5 &   K1III  \\

\hline
\end{tabular}
\end{footnotesize}
\end{center}
\end{table}

\subsection{Solar-like stars}


The first solar-like star observed with WIRE was $\alpha$~Cen (Rigil
Kentaurus; G2V). Preliminary results based on the 50-d light curve
observed in high-cadence mode were reported by Schou \& Buzasi (2001),
who could claim the first clear detection of the characteristic comb
pattern of $p$ modes in the star. This was confirmed in radial velocity by
Bouchy \& Carrier (2001, 2002). Bedding et al.\ (2004) identified 40 modes
from a multisite radial velocity study, and Kjeldsen et al.\ (2005)
constrained the lifetime of the modes to $\tau=2.3^{+1.0}_{-0.6}$\,days.
The main limitation on the uncertainty of the lifetime is the limited time
baseline. Fletcher et al.\ (2006) recognized this, made a refined
analysis of the WIRE data set and measured a mode lifetime of $\tau =
3.9\pm1.4$ days which is in agreement with the result from the radial
velocity survey.

Karoff et al.\ (2007) applied the same method to the WIRE data of the
evolved solar-like star $\beta$~Hydri (G2IV).  They found clear evidence
of solar-like oscillations and measured a mode lifetime very similar to
$\alpha$~Cen ($\tau = 4.2^{+2.0}_{-1.4}$\,d).

Like $\beta$~Hydri, $\alpha$~CMi (Procyon; F5IV-V) is slightly more
massive and more evolved than the Sun. Bruntt et al.\ (2005) found excess
power in the power spectrum which they interpreted as a combination of
granulation and solar-like oscillations. This was in disagreement with the
null result by Matthews et al.\ (2004) based on 32 days of continuous
photometry from the MOST satellite. As discussed by Bruntt et al.\ (2005),
the noise level per data point in the MOST data was more than three times
higher than in the WIRE data. This is likely due to high scattered light
levels in the MOST data (see also Bedding et al.\ 2005).

\subsection{Delta Scuti Stars}

Several \dss\ have been monitored with WIRE. Poretti et al.\ (2002) made
an analysis of the binary \ds\ $\theta^2$~Tau (primary is A7III) and found
12 frequencies which were in agreement with results by Breger et al.\
(2002) from a ground-based multisite campaign.  The detection of a peak at
high frequency seen in both the WIRE and ground-based data lead Breger et
al.\ (2002) to argue that this mode is real (i.e.\ not an alias or
combination frequency) and likely due to oscillations in the secondary
star in the $\theta^2$~Tau binary system.

Poretti et al.\ (2002) were the first to point out that WIRE is capable of
doing time-series of the often neglected brightest stars in the sky, which
are simply too bright for typical 0.5--1.0-m telescopes normally used for
multisite campaigns on \dss\ (e.g.\ the DSN and STEPHI networks).  
Indeed, Buzasi et al.\ (2005) found seven low-amplitude (0.1--0.5 mmag)
modes in $\alpha$~Aql (Altair; A7V), which is now the brightest \ds\ at
$V=0.8$.

Bruntt et al.\ (2007a) combined WIRE photometry and Str\"omgren $uvby$
ground-based observations in an attempt to identify the modes in the \ds\
$\epsilon$~Cep (F0IV). The space-based data provided a superior spectral
window and low noise level.  Using the extracted frequencies from WIRE
they measured the amplitudes and phases in the $uvby$ filters from the
ground-based photometry.  However, the limited amount of ground-based data
made the accuracy of the amplitudes and phases too poor to be able to
identify the modes from phase differences and amplitude ratios (e.g.\
Garrido, Garcia-Lobo \& Rodriguez 1990). Bruntt et al.\ (2007a) estimated
that it would require more than 100 nights of data to obtain the accuracy
on the phases and amplitudes to be able to identify the modes.


\subsection{B-type stars}

More than 35 $\beta$ Cep and SPB stars have been observed with WIRE.
Cuypers et al.\ (2002) confirmed the variability known from spectroscopy
of $\beta$~Cru (Mimosa; B0.5IV)  and in addition found new low-amplitude
modes ($A\simeq0.2-0.3$ mmag).  Cuypers et al.\ (2004) analysed WIRE data
of the known multi-periodic $\beta$~Cep star $\kappa$~Sco (part of Girtab;
B1.5III) and also detected low-amplitude modes not observed previously.

Bruntt \& Buzasi (2006a) gave preliminary results for $\lambda$~Sco
(Shaula; B2IV) which is a known triple system (Uytterhoeven et al.\ 2004).
From spectroscopy it is known that $\lambda$~Sco comprises two B type
stars in a wide orbit ($P\simeq1083$\,d); one of these components has a
low mass companion ($P\simeq5.95$\,d). After subtracting the $\beta$~Cep
pulsation Bruntt \& Buzasi (2006a) could clearly see the primary and
secondary eclipses in the close system. From their preliminary light curve
analysis they constrained the mass and radius of the component stars.

\subsection{Giant stars}

The giant stars comprise around half of the targets observed with WIRE
(cf.\ Fig.~2). This is because only the main target is chosen, while four
additional secondary targets are selected automatically by the on-board
computer based on the apparent brightness of stars in the field of view
(about 8$^\circ$ square).

Buzasi et al.\ (2000) claimed the detection of a comb-like pattern below
25\,\mhz\ ($P>0.5$\,d) associated with solar-like oscillations in
$\alpha$~UMa (Dubhe; K0III). In addition, two significant peaks were found
above the acoustic cut-off frequency (see Dziembowski et al.\ 2001;
Guenther et al.\ 2000). Retter et al.\ (2003) also found a series of peaks
around 4.1\,\mhz\ ($P\simeq2.8$\,d) in $\alpha$~Boo (Arcturus; K1.5III).  
However, their simulations of a pure noise source showed similar spacings
as found in both $\alpha$~UMa and $\alpha$~Boo. The spacings reported in
the two stars are $\Delta \nu=2.9\pm0.3$\,\mhz\ and $\Delta
\nu=0.83\pm0.05$\,\mhz. This is uncomfortably close to the frequency
resolution at $1/T_{\rm obs}=1.1$\,\mhz\ and $0.6$\,\mhz\ for the data
sets of $\alpha$~UMa and $\alpha$~Boo, respectively.

To conclude, the WIRE photometry of K giant stars shows clear evidence of
excess power at low frequencies. In order to investigate whether this is
due to solar-like oscillations and to find further evidence of a comb-like
pattern, a larger sample of bright K~giant stars is currently being
analysed.

\subsection{Eclipsing binary stars}

Bruntt et al.\ (2006b) discovered that $\psi$~Cen (A0IV) is a bright
detached eclipsing binary (dEB), based on photometry from WIRE and the
Solar-Mass Ejection Imager (Howard et al.\ 2006) on the Coriolis
spacecraft. The $\psi$~Cen system comprises a B9 and an A2 type star in an
eccentric orbit ($e=0.55$) with a long period ($P=38.8$\,d). Bruntt et
al.\ (2006b) determined the fractional radii of the stars to just 0.1\%.  
In addition they found evidence of $g$-mode oscillations in the primary
star, despite the star being somewhat cooler than the predicted SPB
instability strip. I am currently analysing spectra of $\psi$~Cen to
determine absolute radii and masses with accuracies better than 0.5\%.

Realizing the unique potential of WIRE to measure masses and radii of
detached dEBs with unprecedented accuracy, a program has been started to
monitor about a dozen known bright eclipsing binaries. Bruntt \&
Southworth (2007) presented preliminary light curves of the known
Algol-type systems AR~Cas (B4IV) and $\beta$ Aur (Menkalinan; A2IV).


\section{Discussion}

I have given an overview of the different classes of stars observed with
the WIRE satellite.  It is interesting that a star tracker never designed
for the purpose has in fact resulted in important discoveries.  One
important lesson learned from WIRE is that accurate pointing (attitude
control) is important when flat fields are not available.  Also, it is of
tremendous value to have the ``raw data'' in the form of individual CCD
windows. With this in hand one can correct for instrumental effects like
scattered light, sub-pixel drift etc.  

In the near future the dedicated photometry missions COROT and Kepler will
provide high precision photometry with much longer time baselines (150\,d
for COROT; up to six years for Kepler)  and nearly 100\% duty cycle.  
This will be particularly interesting for long-period variables and may
potentially solve the ambiguous results from WIRE for the K~giants as was
discussed here. However, less costly small satellites are also being
planned (e.g.\ BRITE; Weiss 2007) and will likely result in interesting
science of bright stars. 

The WIRE results for \dss\ and B-type stars point to the important fact
that detailed comparison with theoretical models is not possible due to
the lack of mode identification.  This must be considered carefully when
planning the ground-based support for the upcoming missions.

\acknowledgments{ It was Derek L.\ Buzasi (US Air Force Academy) who had
the bright idea to use the failed WIRE satellite to do asteroseismology
from space.  I started working with DLB in 2003 and spent five months with
his group at USAFA during 2004. Our collaboration has been very fruitful
as we continue to monitor bright stars with WIRE. I received support from
the Danish Research Agency (Forskningsr\aa det for Natur og Univers), the
Instrument center for Danish Astrophysics (IDA), and the Australian
Research Council. } 

\References{Bedding, T.~R.\ et al.\ 2004, \apj, 614, 380\\

Bedding, T.~R.\ et al.\ 2005, \aap, 432, L43\\

Bouchy, F.\ \& Carrier, F.\ 2001, \aap, 374, L5\\

Bouchy, F.\ \& Carrier, F.\ 2002, \aap, 390, 205\\

Breger et al.\ 2002, \mnras, 336, 249\\

Bruntt, H., Kjeldsen, H., Buzasi, D.~L., \& Bedding, T.~R.\ 2005, \apj,
633, 440\\

Bruntt, H.\ \& Buzasi, D.~L.\ 2006a, Memorie della Societa Astron.\
Italiana, 77, 278\\

Bruntt, H. et al.\ 2006b, \aap, 456, 651\\

Bruntt, H.\ et al.\ 2007a, A\&A, in press, preprint astro-ph/0610539\\

Bruntt, H., Southworth, J.\ 2007b, Proc.\ of IAU~240, preprint
astro-ph/0610540\\

Buzasi, D.\ et al.\ 2000, \apjl, 532, L133\\

Buzasi, D.~L.\ 2001, ASP Conf.~Ser.~223, 389\\

Buzasi, D.\ 2002, ASP Conf.~Ser.~259, 616\\

Buzasi, D.~L.\ 2004, ESA SP-538: Stellar Structure \& Habitable Planet
Finding, 205\\

Buzasi, D.~L.\ et al.\ 2005, \apj, 619, 1072\\

Cuypers, J.\ et al.\ 2002, \aap, 392, 599\\

Cuypers, J., Buzasi, D.\ \& Uytterhoeven, K.\ 2004, ASP Conf.~Ser.~310,
251\\

Fletcher, S.~T.\ et al.\ 2006, \mnras, 371, 935\\

Garrido, R., Garcia-Lobo, E. \& Rodriguez, E.\ 1990, \aap, 234, 262\\

Guenther, D.~B.\ et al.\ 2000, \apjl, 530, L45\\

Hacking, P.\ et al.\ 1999, ASP Conf.~Ser.~177, 409\\

Howard, T.~A. et al.\ 2006, J.\ of Geophys.\ Res.\ (Space Physics), 111,
A04105\\

Karoff, C.\ et al.\ 2007, these proceedings\\

Laher, R.\ et al\, Proc.\ of the 2000 AAS/AIAA Spaceflight Mechanics
Meeting, 146\\

Matthews, J.~M.\ et al.\ 2004, \nat, 430, 51.\ Erratum: 2004, \nat, 430,
921\\

Poretti, E.\ et al.\ 2002, \aap, 382, 157\\

Retter, A.\ et al.\ 2003, \apj, 591, L151.\ Erratum: 2003, \apj, 596,
125\\

Schou, J.\ \& Buzasi, D.~L.\ 2001, ESA SP-464: SOHO 10/GONG 2000, 391\\

Weiss, W.\ W. 2007, these proceedings}

\end{document}